\documentclass{elsart5p}

\usepackage{graphicx}
\usepackage{amssymb}
\usepackage[dvips]{color}

\newcommand{\tl}[1]{\textnormal{\footnotesize #1}}

\begin{document}

\begin{frontmatter}

\title{Unrecognized Backscattering in Low Energy Beta Spectroscopy}

\author{M.~Schumann\corauthref{cor}},
\corauth[cor]{Corresponding author. Tel: +49 6221 549345; Fax: +49 6221 549343}
\ead{marc.schumann@gmx.net}
\author{H.~Abele}
\ead{abele@physi.uni-heidelberg.de}
\address{Physikalisches Institut der Universit\"at Heidelberg, Philosophenweg 12, 69120 Heidelberg, Germany}

\begin{abstract}
We present studies on electron backscattering from the surface of plastic scintillator beta detectors.
By using a setup of two detectors coaxial with a strong external magnetic field -- one detector serving as primary
detector, the other as veto-detector to detect backscattering -- we investigate amount and spectrum of
unrecognized backscattering, i.e.\ events where only one detector recorded a trigger signal. 
The implications are important for low energy particle physics experiments.
\end{abstract}

\end{frontmatter}

% main text
%------------------------------------------------------------------------
\section{Introduction}\label{Introduction}

In the $\beta$-decay of the free neutron in proton, electron,
and anti-neutrino, $n \to p e^- \overline{\nu}_e$,
a number of interesting questions of particle physics and cosmology
can be addressed \cite{Abe07}. These studies provide clean data, and uncertainties
due to nuclear structure do not arise.
Main observables
are correlations (emission asymmetry parameters) between neutron spin and the momenta of the decay products, or 
correlations between two of the latter. They are determined by measuring the emission direction of
electrons and protons (e.g. \cite{Ser98,Abe02,Kre05,Sch07}). Due to their small mass, 
electrons may be scattered from nuclei at large angles leading to electron backscattering out 
of the detector \cite{Leo94}. As generally in low-energy beta spectroscopy,
this induces a systematic effect for neutron decay experiments 
since energy and angle dependent losses may falsify the asymmetry signal \cite{Wie05,Mar06,Wie05b},
a problem that has been discussed for many years now. 
In this paper, we present a method to directly determine the effect.

At present and in the near future, there will be several spectrometers employing a magnetic 
field to precisely measure neutron decay parameters, such as
\mbox{aSPECT}, PERKEO~III, and a PNPI experiment
in Europe, and UCNA, aCORN, abBA, Nab, and PANDA in the U.S. \cite{Abe07,Nic05,NIS05}.
Therefore it is of great interest to study electron backscattering in the framework of
these experiments, to investigate how backscattering can be suppressed, and to determine the
size of possible systematic effects. Especially for plastic scintillators that are widely used 
in these studies almost no data is available in the low energy range \cite{Mar06}. 

\begin{figure}[t]
\begin{center}
\includegraphics*[height=3cm]{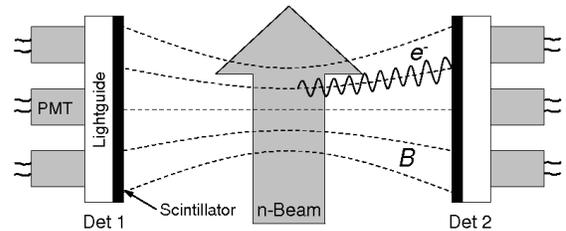}
\caption{The experimental setup of PERKEO II to study electron backscattering. 
Electrons from neutron decay are guided onto the detectors (plastic scintillator) by the magnetic field.}\label{setup}
\end{center}
\end{figure}

%------------------------------------------------------------------------
\section{Experimental Setup}\label{Setup}

We will focus on the rather simple setup of the electron spectro\-meter \mbox{PERKEO II} \cite{Abe97}. This 
is a quite general approach since many of the newly proposed experiments have a similar design. It
features a strong magnetic field \mbox{($B_\tl{max}$ = 1.03 T)} applied across a
polarized neutron beam. The field is slightly decreasing and guides the electrons generated in neutron decay
onto two opposite detectors. In this way, a 2 $\times$ 2 $\pi$ detection system is realized 
and no particles can miss the detectors (Fig.\ \ref{setup}). This configuration is ideally
suited to study backscattering effects from a primary detector which can be registered 
in the secondary detector on the opposite side (``veto detector''). 
The spatially varying magnetic field $B$ also lowers electron backscattering considerably 
due to the magnetic mirror effect: The magnetic flux enclosed by the electron trajectories 
is an adiabatic invariant, 
$p_t^2/B =$ const.,
with the transversal momentum $p_t$. An electron moving in an increasing field can 
therefore be reflected. Many electrons scattered out of the detector are thereby returned 
to the same scintillator where they depose their remaining energy. 
 
The detectors consist each of a large area plastic scintillator (190 $\times$ 130 mm$^2$, 5 mm thick,
Bicron BC 404, pulse width \mbox{2.2 ns}), coupled to a 30 mm thick plexiglass light guide and 
six photomultiplier tubes (Hamamatsu R5504 mesh PMTs). 
These are read out by charge integrating ADCs (analogue to 
digital converter) to measure the energy. A trigger is generated, when at least two 
of the six photomultiplier signals of one detector
have passed the discriminator threshold. The trigger
probability is 90\% at about 100 keV. Every trigger signal is sent to an individual 
TDC (time to digital converter) channel to determine which of the
detectors were hit and to register timing information. A detailed description
of the experimental setup can be found in \cite{Sch07b}.

Compared to solid state detectors, which have backscattering probabilities $p_\tl{BS}$ of
up to 80\% (NaI, \cite{Leo94}), plastic scintillators have the lowest $p_\tl{BS}$ 
in beta spectroscopy due to their low average atomic number $Z$. 
But with  \mbox{$p_\tl{BS} \approx 8\%$} (\mbox{$p_\tl{BS} \approx 4\%$} for normal incident)  
this probability is still quite high \cite{Leo94} and has to be considered in the
analysis of precision measurements.

In the experiments listed above, backscattering enters in an integral way, i.e.\ integrated over
different angles of incidence $\theta$ on the detector. In PERKEO~II, the maximal angle $\theta_\tl{max}$
is around 45$^\circ$ as the decreasing magnetic field increases
the electron momentum component parallel to the field lines. Most electrons hit
the detector near $\theta_\tl{max}$. Overall, the influence of the magnetic field reduces the 
backscattering probability to below 5\%.

%------------------------------------------------------------------------
\section{2-Trigger Backscattering} 

\begin{figure}[t]
\begin{center}
\includegraphics*[width=7.5cm]{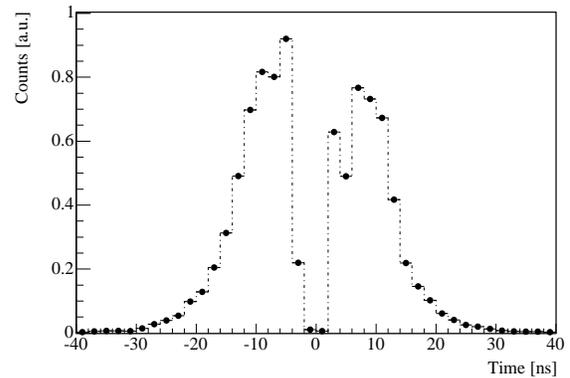}
{\caption{ Timing measurement of 2-trigger backscattering. The plot shows the
time difference between triggers of detector 1 and 2. Events where detector 1 triggered first
are in the left peak. The peaks are well separated.} \label{Fig_Camel} }
\end{center}
\end{figure}

Whenever a trigger signal occurs, the ADCs of 
both detectors are read out simultaneously.
With 180 ns, the integration time is much higher than the average time the backscattered electrons
need to cover the distance between the detectors (800 mm), and we always obtain the full energy
information of the event by summing up both detectors.
\emph{2-trigger backscattering} occurs when an event generates a trigger both in the primary and
the secondary detector. This allows to determine the chronological order
of the two signals by using the timing information of the TDC. If its time resolution is smaller
than the minimal flight time between the detectors, this assignment can be done without
any uncertainty. 

\begin{figure}[tb]
\begin{center}
\includegraphics*[width=7.5cm]{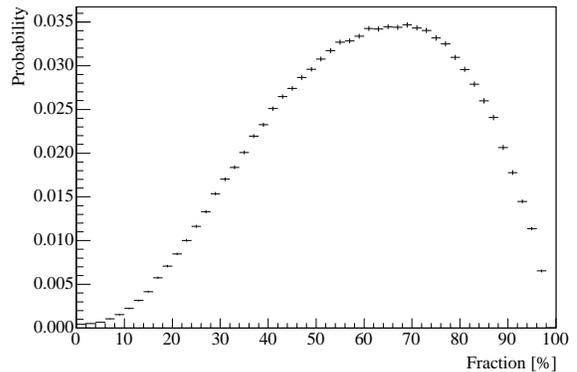}
{\caption{ Measurement of the probability that a backscattering event deposits 
a particular fraction of its energy in the primary detector. The distribution is asymmetric, 
i.e.\ there is a strong preference to depose more energy in the primary detector.} \label{Fig_BSFraction} }
\end{center}
\end{figure}

Fig.\ \ref{Fig_Camel} shows the timing measurement of 2-trigger backscattering: The well
separated peaks correspond to events where detector 1 (left) or detector 2 (right) was hit first.
The separation is a measure of the system's time resolution which is given by the TDC-channel width
of \mbox{0.8 ns}. In between the peaks, where no first detector can be assigned properly, there are less 
than 0.2\% of the backscatter events. Combined with the backscatter probability of below 5\%, this
fraction is negligible.
The energy of the backscattered electrons is not distributed uniformly into primary and secondary
detector (Fig.\ \ref{Fig_BSFraction}): Independent of the
overall signal size, it shows a strong preference to depose more energy in the first.

In summary, 2-trigger backscattering can be fully reconstructed: The events
are assigned to the correct detector with the correct energy.

%------------------------------------------------------------------------
\section{Unrecognized Backscattering and wrongly assigned Events}

\begin{figure}[t]
\begin{center}
\includegraphics*[width=3.5cm, angle=270]{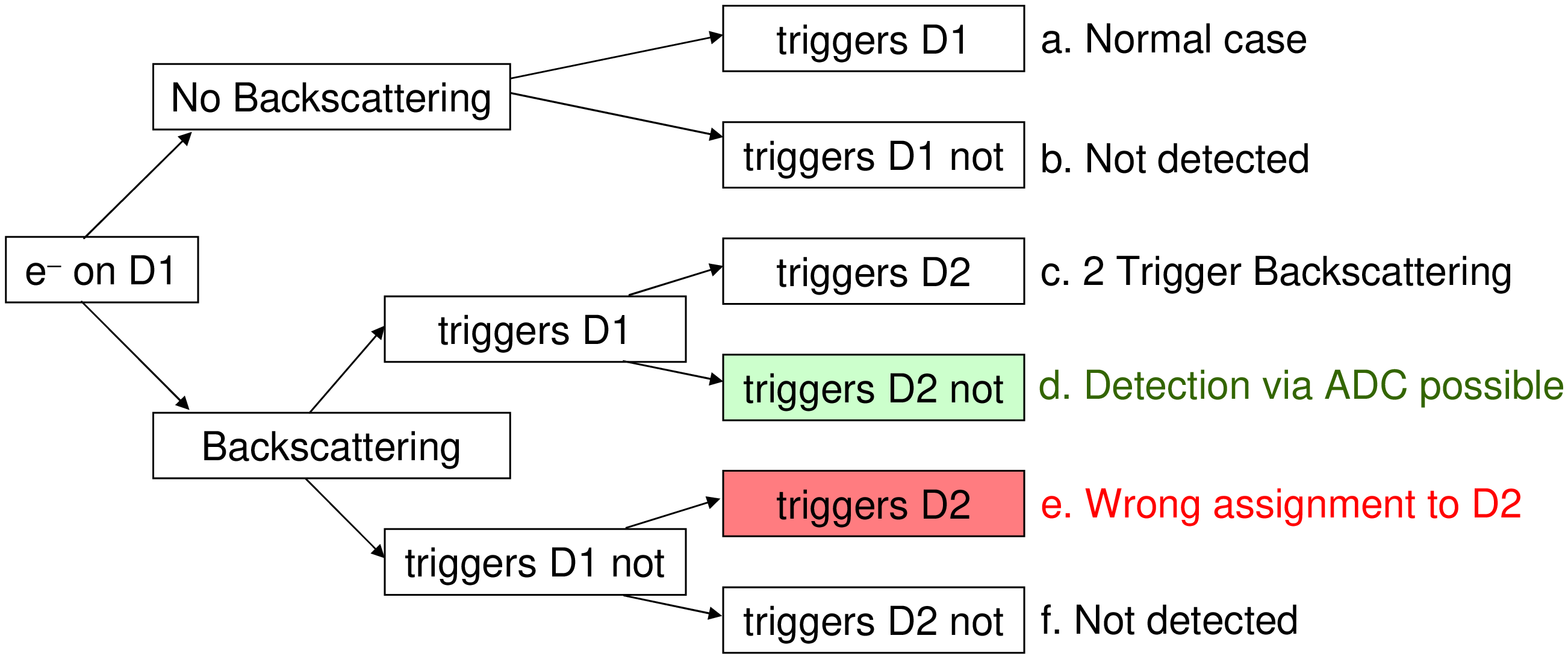}
{\caption{ Backscattering decision tree: The cases $d$ and $e$ 
are important for unrecognized backscattering, but only $e$ 
alters the asymmetry parameter measurements.} \label{Fig_BSTree} }
\end{center}
\end{figure}

In order to analyze the effects of \emph{unrecognized backscattering}, i.e. backscattering where we
do not have two triggers and therefore cannot proceed as described above, we have
to take a look at the decision tree shown in Fig.\ \ref{Fig_BSTree}. We consider an electron
hitting detector 1: If no backscattering occurs, it is not important whether detector 1
triggers or not. In the first case ($a$), we have the usual electron detection and in the second ($b$)
nothing happens at all -- but this case is limited to small energies and is described
by the trigger function of detector 1.

Now consider the case with backscattering. When detector 1 and detector 2 record a trigger (case $c$)
we have 2-trigger backscattering as described above. For small energies it is possible that
both detectors do not trigger and the event gets lost ($f$). When
detector 1 triggers and detector 2 does not, we have unrecognized backscattering, but
the event is assigned to the correct detector ($d$). Crucial is only case $e$, where the primary detector
does not trigger but the secondary does. Here the event is assigned to the
wrong detector -- imposing a systematic error to the asymmetry parameter measurement.
In the following, we will show how to identify unrecognized backscattering and how to 
discriminate between the cases $d$ and $e$.

\begin{figure}[b]
\begin{center}
\includegraphics*[width=7.5cm]{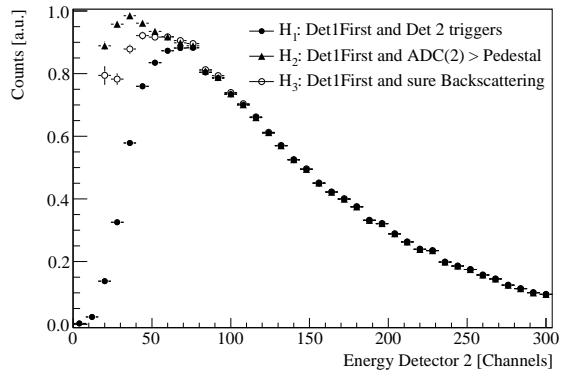}
{\caption{ Lower part of the energy spectrum of the second detector under the
condition that detector 1 has triggered first (Det1First). $H_1$ includes all backscatter 
events with two triggers. The difference between $H_2$ and $H_3$ are the events wrongly assigned 
to detector 1 (cf.\ text).} \label{Fig_BSDetermination} }
\end{center}
\end{figure}

As the energy information is always available for both detectors (even if only one triggers)
we can analyze the ADC content of the second: If the primary detector has triggered and the ADC of
the secondary contains a signal above the pedestal threshold (the ADC signal without any
energy deposition), we have identified an unrecognized backscatter event.
This is illustrated in Fig.\ \ref{Fig_BSDetermination}:
Histogram $H_1$ (solid circles) shows the energy content of detector 2,
when detector 1 has created the primary and detector 2 a secondary trigger (2-trigger 
backscattering, case $c$). $H_2$ (triangles) includes events 
where detector 1 has triggered first and the ADC content of the
second is above the pedestal threshold. For energies high enough to
generate a second trigger, the curves coincide. For low energies, 
$H_2$ shows additional unrecognized backscattering where detector 2 has not
triggered.

The trigger signatures of the cases
\begin{itemize}
\item primary Det1 triggers, secondary Det2 triggers not
\item primary Det2 triggers not, secondary Det1 triggers
\end{itemize}
are the same, whereas the signals belong
to different ``primary'' detectors. Hence we have to discriminate between these cases.

The trigger function gives the probability $T_i(E)$ that an electron of energy $E$
generates a trigger signal in \mbox{detector $i$}. If this probability is unity for the whole
energy range, we would not have any unrecognized backscattering.
Thus we correct the 2-trigger spectrum $H_1$ in Fig.\ \ref{Fig_BSDetermination} to obtain a trigger
probability $T_2(E) = 1$ for all $E$
by dividing it by the measured trigger function of detector 2. The resulting spectrum $H_3$ 
(open circles) contains all events where \mbox{detector 1} has triggered first and the backscattered
electron has reached detector 2.

The difference between the spectra $H_2$ and $H_3$ is the fraction of \emph{wrongly assigned 
events} (case $e$). Here, the experimental signature suggests that 
the events belong to detector 1, 
in reality, however, the electrons have hit detector 2 first
without deposing enough energy to create a trigger.

\section{Size and Spectrum of wrongly assigned Events} 

In this section, we will show that 
the influence of wrongly assigned events (case $e$ in Fig.~\ref{Fig_BSTree})
is negligible in a measurement if the region of
interest starts at energies above a certain threshold. In order to
analyze unrecognized backscattering quantitatively,
it is necessary to extra\-polate the energy spectra in Fig.~\ref{Fig_BSDetermination} 
to lower ADC channels, as it is not possible to evaluate the
spectra below a certain channel due to the pedestal threshold\footnote{Below channel $\sim$16,
the pedestal prevents a correct identification of backscattering events, and the histogram $H_2$
diverges.}. We have chosen two extreme extrapolation cases for the histograms $H_2$ and $H_3$:
In the first we assume no entries to be in the lowest bins, in the second we set them
to the value of the lowest correctly determined channel.

We average over both extrapolation cases to obtain a value for wrongly assigned backscattering,
and choose the difference between average and extrapolation
to be the 2$\sigma$ error. This is a reasonable procedure to account for statistical errors,
extrapolation, and a non-linear detector calib\-ration in this energy region.
Table \ref{results} shows the results: Integration of $H_1$ yields the 2-trigger backscattering, 
the integral of $H_3$
(extrapolated as described above) gives the number of ``true''
backscattering events, where the correct detector has triggered first. The
difference $H_2\!-\!H_3$ gives the fraction of wrongly assigned events: These are
less than 5\% of all backscattering events or $\approx 0.2\%$ of all events.

\begin{table}[t]
\centering
\begin{tabular*}{\columnwidth}{l @{\hspace{0.5cm}}|@{\hspace{0.5cm}} c @{\hspace{0.5cm}} c @{\hspace{0.5cm}} c}
\hline \hline
			 & 2-trigger 	& \ ``true'' BS \qquad & wrong Det \quad \\ \hline
Detector 1 First  	 &  4.23(1)\%	& 4.9(2)\%		& 0.12(1)\% \\
Detector 2 First	 & 3.57(1)\%	& 4.4(2)\%		& 0.20(2)\% \\ \hline \hline
\end{tabular*}
\vspace{0.2cm}
\caption{Results of the general amount of backscattering (``true'' BS) and of the fraction 
assigned to the wrong detector. Since detector 1 had a slightly worse trigger
function and a higher pedestal threshold, its number of wrongly assigned events is
higher.  \label{results}}
\end{table}

\begin{figure}[b]
\begin{center}
\includegraphics*[width=7.5cm]{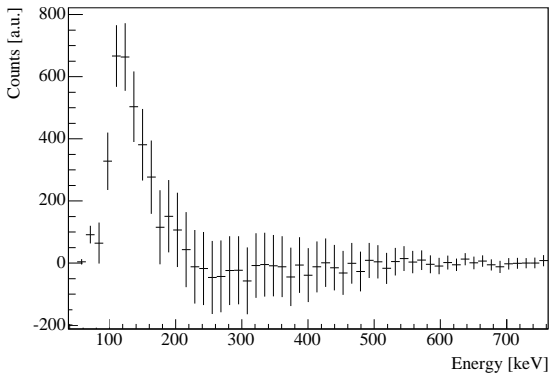}
{\caption{ Total energy of wrongly assigned events: These are
registered as ``detector 1 first'' although detector 2 was hit first without
generating a trigger. The plot shows that unrecognized backscattering is a low energy
effect. Note that the energy scale is uncertain below 100 keV due to
scintillation output non-linearities.} \label{Fig_BSSpectrum} }
\end{center}
\end{figure}

For the analysis of a low-energy experiment it is important to know the energy of
the events assigned to the wrong detector to check their influence.
Fig.\ \ref{Fig_BSSpectrum} shows the full energy spectrum of events in 
$H_2\!-\!H_3$: Within the errors, there are no 
events above \mbox{240 keV}. In general, however, this result depends on the trigger thresholds of the
detectors. We think that this is the first time that size and spectrum of wrongly assigned events 
have been measured in this type of experiment.

\begin{figure}[t]
\begin{center}
\includegraphics*[width=7.5cm]{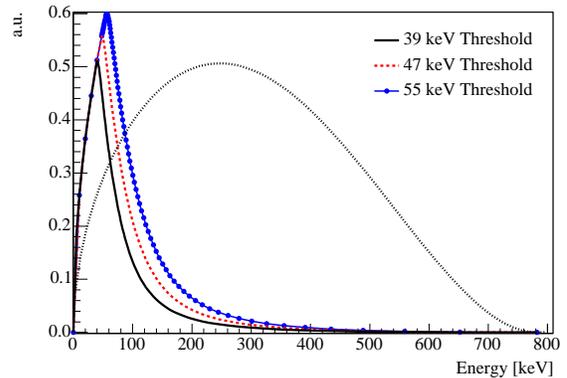}
{\caption{ Modeled energy spectrum of the events wrongly assigned to
the first detector, for different trigger thresholds. The shape is consistent with the
experimental data, Fig.\ \ref{Fig_BSSpectrum}. For comparison, the energy spectrum $F(E)$ of
electrons generated in neutron decay is also given; its endpoint energy is 782 keV.} \label{Fig_BSModel} }
\end{center}
\end{figure}

\begin{figure}[t]
\begin{center}
\includegraphics*[width=7.5cm]{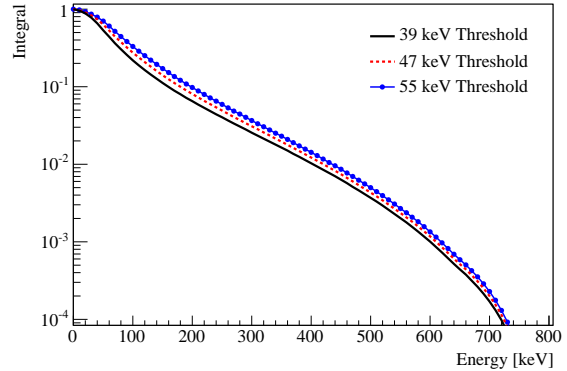}
{\caption{ Integral of the three model spectra of Fig.\ \ref{Fig_BSModel}: 
Above \mbox{200 keV}, the fraction of wrongly assigned backscatter 
events is below \mbox{10\%}.} \label{Fig_BSModelInt} }
\end{center}
\end{figure}

The energy spectrum, Fig.\ \ref{Fig_BSSpectrum}, can also be modeled: 
We start with the normalized distribution $P_{E^*}(E)$, Fig.\ \ref{Fig_BSFraction}, giving the
probability that a backscattered
electron of total energy $E^*$ deposes an energy fraction $E$ in the primary detector. A low
energy threshold (a step-function at $x_0\!=\!39$, 47, \mbox{and 55 keV}) is introduced to account for the
trigger function of
the first detector. The distribution is integrated for different electron energies $E^*$ from 0 to $x_0$
to determine the fraction of events not triggering the first detector. This is then 
multiplied with the Fermi-spectrum $F(E^*)$ 
to obtain an energy distribution similar to the situation in neutron decay. 
The model yields the spectrum
$H_2\!-\!H_3$:
\begin{equation}
s(E^*) = F(E^*) \ \int_{0}^{x_0} P_{E^*} (E) \ \textnormal{d}E  \textnormal{.}
\end{equation}
We neglect the trigger function of the secondary detector.

The model, Fig.\ \ref{Fig_BSModel}, agrees well with the measurement, \mbox{Fig.\ \ref{Fig_BSSpectrum}}.
Integration allows to estimate the fraction of wrongly assigned
events in the spectra, Fig.\ \ref{Fig_BSModelInt}: If the region of interest
starts above 240 keV, the model predicts less than \mbox{6\%} (\mbox{$x_0\!=\!55$ keV}) of wrongly assigned events
to have higher energies. With overall 0.2\% wrongly assigned events \mbox{(Table \ref{results})}, this yields 
a maximal correction in the order of \mbox{0.01\%} which is 
negligible in all ongoing experiments. This result is consistent with the measurement shown in Fig.\ \ref{Fig_BSSpectrum} 
with no wrongly assigned events above 240 keV within the errors.

\section{Conclusion}

In this paper, we have presented a method ­- based on measured data only -­ to analyze electron
backscattering quantitatively. Electrons backscattered from detector surfaces might
seriously affect beta spectroscopy. In a setup of two detectors coaxial with a magnetic field,
electrons backscattered from one detector are guided to the opposite (``veto'') detector and
backscattering effects are well suppressed. In this setup, it is still possible that an
event generates only a trigger in the veto detector and is therefore assigned to the wrong
emission direction, which might give rise to systematic effects. We have
shown that these effects only affect low energies. They are negligible when analyses of the 
electron spectra are performed only above an energy threshold. However, when the low energetic 
part of the spectrum shall be used as well, the effect of wrongly assigned events must be 
corrected using the method discussed in this paper.

\section*{Acknowledgments} 

This work was funded by the German Federal Ministry for Research and Education under
contract number 06HD153I.

\end{document}